\documentstyle[prd,aps,preprint,tighten]{revtex}

\begin{document}
\draft \title{Is there a mass dependence in the spin structure 
of baryons?}

\author{Johan Linde\footnote{E-mail: jl@theophys.kth.se} and H{\aa}kan 
Snellman\footnote{E-mail: snell@theophys.kth.se}} \address{Department 
of Theoretical Physics\\
Royal Institute of Technology\\
S-100 44 STOCKHOLM\\
SWEDEN}

\maketitle
\begin{abstract}
We analyze the axial-vector form factors of the nucleon-hyperon system 
in a model with SU(3)$_{\text{flavor}}$ symmetry breaking due to mass 
dependent quark spin polarizations.  This mass dependence is deduced 
from an analysis of magnetic moment data, and implies that the spin 
contributions from the quarks to a baryon decrease with the mass of 
the baryon.  When applied to the axial-vector form factors, these mass 
dependent spin polarizations bring the various sum-rules from the 
quark model in better agreement with experimental data.  As a 
consequence our analysis leads to a reduced value for the total spin 
polarization of the proton.
\end{abstract}


\narrowtext

\section{Introduction}
In the quark model the axial-vector current form factors measured in 
nucleon-hyperon weak decays are related to the spin polarizations of 
quarks.  In principle there are three quark spin polarization 
parameters, $\Delta u$, $\Delta d$ and $\Delta s$ that can be 
determined from the experimental data.  Unfortunately the four 
axial-vector form factors can all be expressed in term of two 
parameters, $a_{3} =\Delta u-\Delta d$ and $a_{8}=\Delta u +\Delta d - 
2\Delta s$.  To determine the quark spin polarizations we therefore 
also need the total spin polarization $\Delta \Sigma = \Delta u + 
\Delta d + \Delta s$ of the nucleon.

The relation between the axial-vector currents and the nucleon spin has been 
discussed in the recent literature. 

McCarthy {\it et al\/}\cite{McCarty} have 
performed an analysis of the recent 
deep inelastic scattering measurements to obtain the quark spin 
content and concluded that in view of current limited understanding of 
the nature and magnitude of SU(3)$_{\text{flavor}}$ symmetry 
breaking, the use of SU(3) underestimates the 
uncertainties of the flavor-specific quark helicity contributions to 
the nucleon spin.  In another paper \cite{Song} Song {\it et al\/} have 
discussed small SU(3) breaking contributions to the analysis.

Also Lipkin \cite{Lipkin} and Lichtenstadt and Lipkin 
\cite{Lichtenstadt} have analyzed possible SU(3) symmetry breaking in 
a model to estimate its influence on the value of the axial vector 
form factors and its effect on the spin polarization analysis.  Their 
conclusion is that SU(3) symmetry breaking does not have a large 
influence on the nucleon spin evaluation.

Ehrnsperger and Sch{\"a}fer \cite{Schafer} have also performed an 
analysis of hyperon beta decay to extract the $F/D$ ratio and study 
its implication on the spin polarization sum rules.  They argue that 
there is no evidence for a strong spin polarization of the strange 
quark sea.  

Ratcliffe \cite{Ratcliffe} has analyzed hyperon semi-leptonic decays 
and come to the conclusion that SU(3) symmetry with some small 
corrections for its violation gives a very satisfactory description 
of the experimental data.

In all these analyses, both with and without SU(3) symmetry breaking, 
the authors have used the Review of Particle Properties (RPP) data 
set\cite{data} for the hyperon decays, which are obtained by assuming 
that the SU(3) symmetry breaking induced form factors $g_{2}$ can be 
neglected.
 
In our opinion an analysis with SU(3) breaking to evaluate the 
axial-vector form factors should use the best experimental results 
available in the literature, including the symmetry breaking form 
factor $g_{2}$.  This is the attitude adopted in the present work and 
elaborated upon later.  Using the RPP data set already presumes that 
the symmetry breaking effects are small.  This is clearly undesirable 
in an analysis of such effects.

In two earlier papers\cite{jlhs,jlhs2} we have discussed the magnetic 
moments of baryons in an extension of the quark model, which allows 
for general flavor symmetry breaking and where the quark magnetic 
moments are allowed to vary with the isomultiplet $B$.  The magnetic 
moments of the baryons in this model can be written as a linear sum of 
contributions from the various flavors
 \begin{equation}
	\mu(B^{i}) = \mu_{u}^{B}\Delta u^{B^{i}} + \mu_{d}^{B}\Delta 
	d^{B^{i}} + \mu_{s}^{B}\Delta s^{B^{i}},
	\label{moment}
\end{equation}
where $\mu_{f}^{B}$ is the effective magnetic moment of the quark of 
flavor $f$ in the isomultiplet $B$ and $\Delta f^{B^{i}}$ is the 
corresponding spin polarization for baryon $B^{i}$, $i$ being the 
baryon charge state.  By symmetry arguments the $\Delta f^{B^{i}}$'s 
in the octet baryons can be expressed as constant linear combinations 
of the three $\Delta f$'s for the proton, which are the only spin 
polarizations needed to describe the octet:
\begin{equation}
	\Delta f^{B^{i}} = \sum_{f'}M(B^{i})_{ff'}\Delta f',
	\label{deltabi}
\end{equation}
where $f,f'$ runs over $u,d,s$, and the $M(B^{i})$'s are matrices with 
constant elements.  In particular for the six mirror symmetric baryons 
of type $B(xyy)$, where $x$ and $y$ are different flavors, we have 
$\Delta y^{B^{i}}=\Delta u$, $\Delta x^{B^{i}}=\Delta d$ and $\Delta 
z^{B^{i}}=\Delta s$ where the flavor $z$ is the non-valence quark 
flavor.  

In the non-relativistic quark model (NQM) the values of the spin 
polarizations are $\Delta u = \frac{4}{3}$, $\Delta d = -\frac{1}{3}$ 
and $\Delta s = 0.$

Due to the homogeneity of the right hand side of (\ref{moment}), it is 
a question of definition if the dependence on the baryon multiplet is 
considered to be associated with the quark magnetic moment rather than 
with the spin polarization.  In Refs.\cite{jlhs} and \cite{jlhs2} we 
have chosen to analyze the data by keeping the spin polarizations 
fixed throughout.

Here we will analyze the opposite situation, where the spin 
polarization is instead assumed to vary with the baryon multiplet and 
the quark magnetic moments are the same for all multiplets.

This scheme has the advantage of making the properties of the quarks 
static and environment independent.  Since the effective magnetic 
moment of a quark in the NQM has the form
\begin{equation}
	\mu_{f}=\frac{e_{f}}{2m_{f}},
\end{equation}
$e_{f}$ being the quark charge, this means that there is no dependence 
of the effective quark mass $m_{f}$ on $B$.  This is in accordance 
with the fact that the same constituent quark masses can be used 
successfully to predict the baryon octet and decuplet masses with only 
a hyperfine splitting interaction in the Hamiltonian.  The 
disadvantage is that the spin structure varies from multiplet to 
multiplet. 

This variation of spin polarization with the isospin multiplet introduces 
an explicit SU(3)$_{{\rm flavor}}$ symmetry breaking in the analysis, which 
is different from earlier symmetry breaking schemes and is evaluated
from the measured baryon magnetic moments. 

A further merit of this interpretation, and the one 
that we are going to analyze here, is that the sum-rules governing the 
axial-vector form factors are better fulfilled in this scheme, 
although the errors are still somewhat large to definitely decide 
between either of the two ways of attributing the mass dependence 
effect.

Since we are going to allow for SU(3) symmetry breaking in our analysis, 
we will use the experimental data including the symmetry breaking 
form factors $g_{2}$ whenever available.

\section{Calculating the model parameters}
The spin structure parameters in the expressions for the magnetic 
moments and in the deep inelastic scattering experiments and 
axial-vector form factors are not a priori the same.  In many models 
they are nevertheless proportional\cite{karl}, and can be normalized 
to be the same.  We normalize them to the axial-vector form factor 
$g_A^{np}=1.2573$, as is generally done.

We write the baryon magnetic moments as
\begin{equation}
	\mu(B^{i}) = \sum_{f,f'}\mu_{f}\alpha(B) M(B^{i})_{ff'}\Delta 
	f' ,
\end{equation}
where $\mu_{f}$ is the magnetic moment of the quark of flavor $f$ and 
$\Delta f$ is the corresponding spin polarization.  The factor 
$\alpha(B)$ is an overall factor, the same for all flavors, depending 
only on the isomultiplet $B$.  The flavor symmetry breaking within the 
isomultiplets is then accounted for by the quark magnetic moments that 
are free parameters.  We stress that the value of $\alpha(B)$ is 
independent of the isospin symmetry breaking parameter 
$T=\mu_{u}^{B}/\mu_{d}^{B}$ discussed in Ref.\cite{jlhs}.  Thus we 
can, if we like, assume that we have isospin symmetry.

In our previous analysis we associated the factor $\alpha(B)$ with the 
quark magnetic moments and defined $\mu_f^{B}=\alpha(B)\mu_f$ as in 
equation (\ref{moment}).  We can choose to normalize $\alpha(B)$ to 
$\alpha(N)=1$, in which case $\mu_{f}^{N}=\mu_{f}$.  The other values 
of $\alpha$ can then be obtained from the previously extracted values 
of the $\mu_{d}^{B}$'s as\cite{jlhs,jlhs2}
\begin{eqnarray}
	\alpha(\Lambda) & = & \mu_{d}^{\Lambda}/\mu_{d}^{N} = 0.88\pm 
	0.04,\label{alphaL}\\
	\alpha(\Sigma) & = & \mu_{d}^{\Sigma}/\mu_{d}^{N} = 0.91\pm 0.01, 
	\\
	\alpha(\Xi) & = & \mu_{d}^{\Xi}/\mu_{d}^{N} = 0.85\pm 0.03.
	\label{alphaX}
\end{eqnarray}
We will now instead associate $\alpha(B)$ with the spin polarizations.  
Equation (\ref{deltabi}) is then rewritten as
\begin{equation}
	\Delta f^{B^{i}} = \sum_{f'}M(B^{i})_{ff'}\alpha(B)\Delta f'.
	\label{nydeltabi}
\end{equation}
Thus, {\it e.g.\/} in the mirror symmetric baryons $B(xyy)$, we 
instead have $\Delta y^{B^{i}}=\alpha(B)\Delta u$, $\Delta 
x^{B^{i}}=\alpha(B)\Delta d$ and $\Delta z^{B^{i}}=\alpha(B)\Delta s$.

The values of $\alpha(B)$ can be well fitted to a linear function of 
the mean mass of $B$ as shown in Fig.~\ref{cool_plot}.  The linear 
relation is
\begin{equation}
	\alpha(m)= 1-0.376(m - 0.939),
	\label{lin}
\end{equation}
when $m$ is expressed in GeV. We will continue to use 
$\alpha(B)\equiv\alpha(m_B)$ in the following.

We will test this linear relation in what follows by 
using the interpolated $\alpha$'s from the equation above.  These 
values are
\begin{eqnarray}
	\alpha(\Lambda) & = & 0.93\pm 0.02, \\
	\alpha(\Sigma) & = & 0.90\pm 0.02 , \\
	\alpha(\Xi) & = & 0.86\pm 0.02.
\end{eqnarray}
To illustrate why this $B$ dependent factor is needed we regard the 
sum-rule\cite{franklin}
\begin{equation}
	\mu(p) +\mu(\Xi^{0}) +\mu(\Sigma^{-}) - \mu(n) - \mu(\Xi^{-}) - 
	\mu(\Sigma^{+}) = 0,
\end{equation}
which follows when the quark magnetic moments and spin polarizations 
both are independent of $B$.  It is badly broken by the experimental 
data by ten standard deviations so that the left hand side is instead 
$0.49\pm 0.05$.  In our more general parameterization this sum-rule 
becomes
\begin{equation}
	 \mu(p) +\frac{\mu(\Xi^{0})}{\alpha(\Xi)} + 
	 \frac{\mu(\Sigma^{-})}{\alpha(\Sigma)} - \mu(n) - 
	 \frac{\mu(\Xi^{-})}{\alpha(\Xi)} - 
	 \frac{\mu(\Sigma^{+})}{\alpha(\Sigma)} = 0.
 \end{equation}
Due to the construction of the $\alpha$'s this sum-rule is satisfied.

There is no doubt that the correction factors are needed.  However, if 
a totally different structure is envisaged with more degrees of 
freedom, like pion-clouds etc.\ there is of course room for other 
possibilities.  Our aim is to investigate the symmetry-breaking in the 
more restricted framework of magnetic moments, spin polarizations, and 
axial-vector form factors, since these are extensively used to 
investigate baryon structure.

\section{The axial-vector form factors}
Whether we associate the $\alpha(B)$ factors to the quark magnetic 
moments or the spin polarizations, obviously does not affect the 
analysis of the magnetic moments.  However, the analysis of the 
axial-vector form factors will be modified when we let the spin 
polarizations be given by (\ref{nydeltabi}).

The axial-vector form factors can in this parameterization now be 
written
\begin{eqnarray}
	g_{A}^{np} & = & \Delta u - \Delta d , \\
	g_{A}^{\Lambda p} & = & \frac{1}{3}(2\Delta u - \Delta d - \Delta 
	s)\alpha(\Lambda) , \\
	g_{A}^{\Xi \Lambda} & = & \frac{1}{3}(\Delta u + \Delta d - 
	2\Delta s)\alpha(\Xi) ,\\
	g_{A}^{\Sigma n} & = &(\Delta d - \Delta s)\alpha(\Sigma) .
\end{eqnarray}
This can be used to derive the two sum-rules
\begin{eqnarray}
		\frac{g_{A}^{\Xi \Lambda}}{\alpha(\Xi)} + \frac{g_{A}^{\Lambda 
		p}}{\alpha(\Lambda)} & = & \frac{g_{A}^{\Sigma 
		n}}{\alpha(\Sigma)} + g_{A}^{np}, \\
	\frac{g_{A}^{\Xi \Lambda}}{\alpha(\Xi)} + g_{A}^{np} & = & 2 
	\frac{g_{A}^{\Lambda p}}{\alpha(\Lambda)},
\end{eqnarray}
As mentioned in the introduction, since $\alpha(m)$ is an SU(3) breaking 
parameter, it would in our opinion be 
inconsistent to use $g_{A}$'s extracted from experimental data in the 
SU(3) symmetric limit.  In the RPP 
data table\cite{data} the values are effective values of the $g_{A}$'s 
in the SU(3) symmetric limit where the induced axial vector form 
factors $g_{2}$ are taken to be zero.  We want here to explore the 
consequence of taking the symmetry breaking into account through the 
$g_{2}$'s.  The values of the $g_{A}$'s are in general sensitive 
to the values of the $g_{2}$'s\cite{roos}.

For $g_{A}^{\Sigma n}$ we use the value $g_{A}^{\Sigma n}=-0.20\pm 
0.08$ from Hsueh {\it et al.\/}\cite{hsueh} and for $g_{A}^{\Lambda 
p}$ the value $g_{A}^{\Lambda p}=0.731\pm 0.016$ from Dworkin {\it et 
al.\/}\cite{dworkin}.

The value $g_{A}^{\Sigma n}=-0.340\pm 0.017$ in RPP assumes that the 
induced form factor $g_{2}=0$ (SU(3) limit).  The value measured by 
Hsueh {\it et al.\/} is $g_{2}=0.56\pm 0.37$ which is indeed far from 
zero.

Similarly, we use the value for $g_{A}^{\Lambda p}$ from Dworkin {\it 
et al.\/} because they have not assumed that the weak magnetism 
coupling $g_{W}=0.97$.  However, it would not make any significant 
difference to use the value $g_{A}^{\Lambda p}=0.718\pm 0.015$ given 
in the RPP data table.

For $g_{A}^{\Xi\Lambda}$ no value exists with $g_{2}\neq 0$, and 
therefore we take its value directly from the RPP data table. 

The two sum-rules are barely satisfied without the $\alpha$'s.  The 
relations are satisfied as follows
\begin{eqnarray}
(0.98\pm 0.07)\quad 1.07\pm 0.06 & = & 1.04\pm 0.09 \quad (1.06 \pm 
0.08), \\
(1.51\pm 0.05)\quad 1.55\pm 0.06 & = & 1.57\pm 0.05 \quad (1.46 \pm 
0.03),
\end{eqnarray}
corresponding to the two equations above.  The numbers in parentheses 
are the values without the $\alpha$'s (i.e.  $\alpha \equiv 1$).

Although the improvement relative to the case without the $\alpha$'s 
might not be dramatic, both sum-rules are definitely better satisfied 
with the $\alpha$'s.

As a further test we calculate the constant $R=\frac{\Delta u-\Delta 
d}{\Delta u-\Delta s}$ defined in Ref.\cite{jlhs}.  This constant has 
the value $R=1.18\pm0.01$ from the magnetic moment data.  Our 
expression for this constant, expressed in terms of axial-vector form 
factors, is now
\begin{equation}
	R=\frac{2g_{A}^{np}}{ g_{A}^{\Lambda p}/\alpha(\Lambda) + 
	g_{A}^{\Xi \Lambda}/\alpha(\Xi)+ g_{A}^{\Sigma 
	n}/\alpha(\Sigma)+g_{A}^{np}} = 1.19\pm0.06.
\end{equation}
This is again an improvement over the value $R=1.23\pm0.06$ found in 
Ref.\cite{jlhs}.

The four axial-vector form factors can be parameterized by two 
variables, which we choose as $\Delta u- \Delta d=g_A^{np}$ and 
$a_8=\Delta u +\Delta d-2\Delta s$.
 
Since $g_A^{np}=1.2573\pm 0.0028$ is by far the best measured 
parameter we will use this as a fix parameter and express the three 
other axial-vector form factors in terms of $g_A^{np}$ and $a_8$.  
This gives
 \begin{eqnarray}
 \frac{ g_A^{\Lambda p}}{\alpha(\Lambda)} & = & \frac{1}{6}a_8 + 
 \frac{1}{2}g_A^{np}, \label{anpassa1}\\
 \frac{ g_A^{\Sigma n}}{ \alpha (\Sigma)}& = & \frac{1}{2}a_8 
 -\frac{1}{2}g_A^{np} , \label{anpassa2}\\
	\frac{g_A^{\Xi\Lambda}}{\alpha(\Xi)}& = & 
	\frac{1}{3}a_8.\label{anpassa3}
 \end{eqnarray}
 
We have performed two least square fits of $a_8$ using these formulas 
and the experimental numbers quoted above, one with the $\alpha$'s and 
one without.  With the $\alpha$'s we get $a_8=0.89\pm0.08$ with 
$\chi^{2}=0.31$ and without the $\alpha$'s we get $a_8=0.70\pm 0.08$ 
with $\chi^{2}=1.9$.  We see that there is a considerable improvement 
when the $\alpha$'s are included.  This is further illustrated in 
Fig.\ref{cooler_plot}.  The left picture illustrates the fit to Eqs.\ 
(\ref{anpassa1})-(\ref{anpassa3}), without the $\alpha$ factors.  The 
gray bands are the experimental values of the $g_{A}$'s including the 
errors.  The linear functions are the corresponding right hand sides 
of (\ref{anpassa1})-(\ref{anpassa3}).  The thick central vertical line 
is the value for $a_{8}$ obtained in the fit, and the thinner  
vertical lines are the  error bounds of the fit.  The right 
picture similarly illustrates the fit when the $\alpha$'s are 
included.  We can see that all $g_{A}$'s are better accounted for 
when using the $\alpha$ factors.

The value $a_{8}=0.89\pm0.08$ differs rather much from the 
value $a_{8}=0.601\pm 0.038$ used by Ellis and Karliner\cite{Ellis} in 
their proton spin polarization analysis.

The difference has two sources: the $\alpha$ factors and the different 
choice of experimental data for the $g_{A}$'s.  This shows, in our 
opinion, that there 
is a rather large uncertainty in the determination of the quark spin 
polarization of the nucleon, which involves the constant $a_{8}$.  The 
conventional wisdom is to use the experimental values with the 
assumption that $g_{2}=0$ for all $g_{A}$'s.  Using these values in 
Eqs.  (\ref{anpassa1})-(\ref{anpassa3}) with $\alpha\equiv 1$ leads 
to a value of $a_{8} = 0.58 \pm 0.06$. However, such an 
analysis ignores the experimental 
fact that $g_{2}\neq 0$ for $g_{A}^{\Sigma n}$, making the 
determination uncertain. On the other hand, in our analysis we have 
only set $g_{2}\neq 0$ for one $g_{A}$ as the others are 
experimentally unknown. This obviously also leads to an uncertainty in the 
determination.

There are basically three different ways that the data can be 
analyzed at present:

a) We ignore the non-zero value of $g_{2}$ for $g_{A}^{\Sigma n}$ and hope
that there is no effect in the other form factors. In 
this case $\alpha\equiv 1$.

b) We include the correction $g_{2}$ for $g_{A}^{\Sigma n}$ and hope 
that this correction is small for the 
other $g_{A}$'s. $(\alpha\equiv 1.)$

c) We include the correction for $g_{2}$ that is measured, and hope
that they are not too large for the unmeasured ones. We also include
the phenomenological SU(3) correction factors from the  magnetic moment
analysis. $(\alpha\not=1.)$

The option c) is in our view essentially on the same level of
rigor as the option a), which tries to ignore the problem entirely.
We cannot find that it is more consistent to try to ignore a problem
that has been experimentally shown to exist, rather than to treat it,
even if the treatment is incomplete, due to lack of data.

Our calculation has shown that the sum-rules for the $g_{A}$'s seem to 
be better accounted for using the approach c). This indicates 
that the corrections are important.  

We can also see that our approach is more consistent with the 
sum-rules if we calculate $a_{8}$ separately from Eqs.\ 
(\ref{anpassa1})-(\ref{anpassa2}) and (\ref{anpassa3}). Option a) 
gives $a_{8}=0.57\pm0.03$ from (\ref{anpassa1})-(\ref{anpassa2}) and 
$a_{8}=0.75\pm0.15$ from (\ref{anpassa3}). Option c) gives the values 
$a_{8}=0.90\pm0.06$ and $a_{8}=0.87\pm0.13$ respectively. These two 
values are certainly much closer to each other.

\section{Implications for the proton spin polarization analysis}
The values of $g_{A}^{np}$ and $a_{8}$ are commonly used in analyses 
to determine the proton spin polarization.  A change in the value of 
$a_8$ therefore has a non-negligible influence on such a spin 
polarization analysis.  We will illustrate this using the formulas 
from the analysis of Ellis and Karliner\cite{Ellis}.  Their evaluation 
of $\Delta\Sigma=0.31\pm 0.07$ can be expressed as
 \begin{equation}
	\Delta\Sigma(Q^2) = 9\frac{\Gamma_1^p(Q^2) 
	-(\frac{g_A^{np}}{12}+\frac{a_8}{36})f(\alpha_s)}{h(\alpha_s)},
 \end{equation}
 where $f(\alpha_s)$ and $h(\alpha_s)$  are as in \cite {Ellis}. 
The constant $g_A^{np}$ has the usual value $g_A^{np}=1.2573$, but the 
constant $a_8$ has in Ref.\cite{Ellis} the value $a_8=0.601\pm 0.038$.

The value of $\Delta\Sigma$ will change with the value of $a_8$.  Let 
the change in $a_8$ be denoted $\delta a_8$, and 
the new value of $\Delta\Sigma$ be denoted $\Delta\Sigma'$.  We then 
have
 \begin{equation}
	\Delta\Sigma'=\Delta\Sigma-\frac{\delta a_8}{4}\left(1-{\cal O}(2) 
	\left(\frac{\alpha_s(Q^2)}{\pi}\right)^2 \right)\approx 
	\Delta\Sigma-\frac{\delta a_8}{4}.
 \end{equation}
The value of $a_8 =0.89 \pm 0.08$ found above will thus lead to a 
different estimate of the total spin polarization of the proton.  
$\Delta\Sigma$ will change to
 \begin{equation}
	\Delta\Sigma'=0.31- \frac{0.29}{4}=0.24\pm0.09.
 \end{equation}
In our previous analysis of isospin symmetry breaking in the baryon 
magnetic moments\cite{jlhs} this value favors a slightly smaller 
isospin symmetry breaking than the value $\Delta\Sigma=0.31$.  It also 
changes slightly the quark spin content of the proton to the values
 \begin{eqnarray}
	\Delta u & = & 0.86\pm 0.04 , \\
	\Delta d & = & -0.40\pm 0.04 ,\\
	\Delta s & = & -0.22\pm 0.05.
 \end{eqnarray}
calculated by means of $\Delta\Sigma'$, $a_8$ and $g_A^{np}$.  The 
main effect is to allow $\Delta s$ to be larger.

\section{Discussion and conclusions}
As we have seen above there is support from the axial-vector form 
factor data, when $g_{2}$ is taken into account, that the spin 
polarizations of the quarks are diminishing with the increase of mass 
of the host particle.  This mass dependence is born out in the 
sum-rules that can be written and are well satisfied by the 
experimental data. These mass dependent spin polarizations also 
account for the breaking of the sum-rule for the 
baryon magnetic moments.

Our analysis leads to a change in the evaluation of the 
axial-vector coupling constant $a_{8}$, that will affect the proton spin 
polarization analysis.  Our value for the constant $a_8$ favors a 
lower value of the proton quark spin sum.

The data for the axial-vector form factors used in the sum-rules are 
subject to statistical and systematic errors. There are 
essentially three form factor values used by us in these sum-rules, that 
are still uncertain. These are $g_{A}^{\Xi \Lambda}$, $g_{A}^{\Sigma n}$, and 
$g_{A}^{\Lambda p}$. For the first one, which is a low statistics 
experiment (1992 events), the quoted error is the combined 
error. The two last ones are high statistics experiments. 
For the $g_{A}^{\Sigma n}$ form factor the quoted error is probably 
dominated by the error in the determination of the induced form 
factor $g_{2}$. For the $g_{A}^{\Lambda p}$ form factor measured by 
Dworkin et al, there might be a further systematic error $\pm 0.012$ to be 
added to the quoted error. If this is done the difference between the 
two fits in Fig 2. 
becomes somewhat less pronounced, but the main features remain.

The final result for $a_{8}$ should of course therefore be 
taken with some caution 
and await further measurements of the other $g_{2}$ factors to be 
complete.  However, it does, in our opinion, merit consideration and 
shows the uncertainty at present in the determination of $a_{8}$, which 
has a non-negligible influence on the determination of the nucleon quark 
spin polarizations.

The studies performed by Leinweber {\it et al.\/}\cite{draper} 
supports indirectly the findings here.  Their lattice gauge 
calculations have been done in quenched QCD, and to define the 
effective quark magnetic moments, the spin polarizations were set to 
their NQM values once the results were extracted from the lattice.  
The quark magnetic moments then show a decrease in value with 
increasing mass of the host particle, in much the same way as we found 
in Ref.\cite{jlhs2}, where we also chose the keep the spin 
polarizations mass independent.

One possible interpretation of this effect could be that the quarks 
simply gradually lose their orientation as the excitation energy 
increases.

As the total angular momentum of the proton is fixed to $1/2$, this 
means that there must be a contribution from some other electrically 
neutral component that increases its angular momentum with baryon mass 
to compensate for the decrease in the contribution coming from the 
quarks.
 
One possibility is to attribute such a contribution to the presence of 
gluonic components in the baryons.  This is perfectly consistent with 
the findings from deep inelastic scattering experiments, that only 
about half of the proton momentum is carried by the quarks.  Also a 
collective mode of the Skyrmion type, with a rather small contribution 
to the magnetic moment, could be envisaged to manifest in this way.
 
The new feature found here is that this contribution seems to vary 
linearly with the mass of the baryon multiplet.

This emphasizes the importance of trying to measure the magnetic 
moments of high mass baryon states and also to try to calculate them 
with lattice gauge techniques.
 
Also lattice gauge calculations of the axial-vector form factors for 
the heavier states, would possibly shed light on the behavior found 
here.  Such calculations have already been performed for the nucleon 
system\cite{axial}.
 
\acknowledgements This work was supported by the Swedish Natural 
Science Research Council (NFR), contract F-AA/FU03281-308.

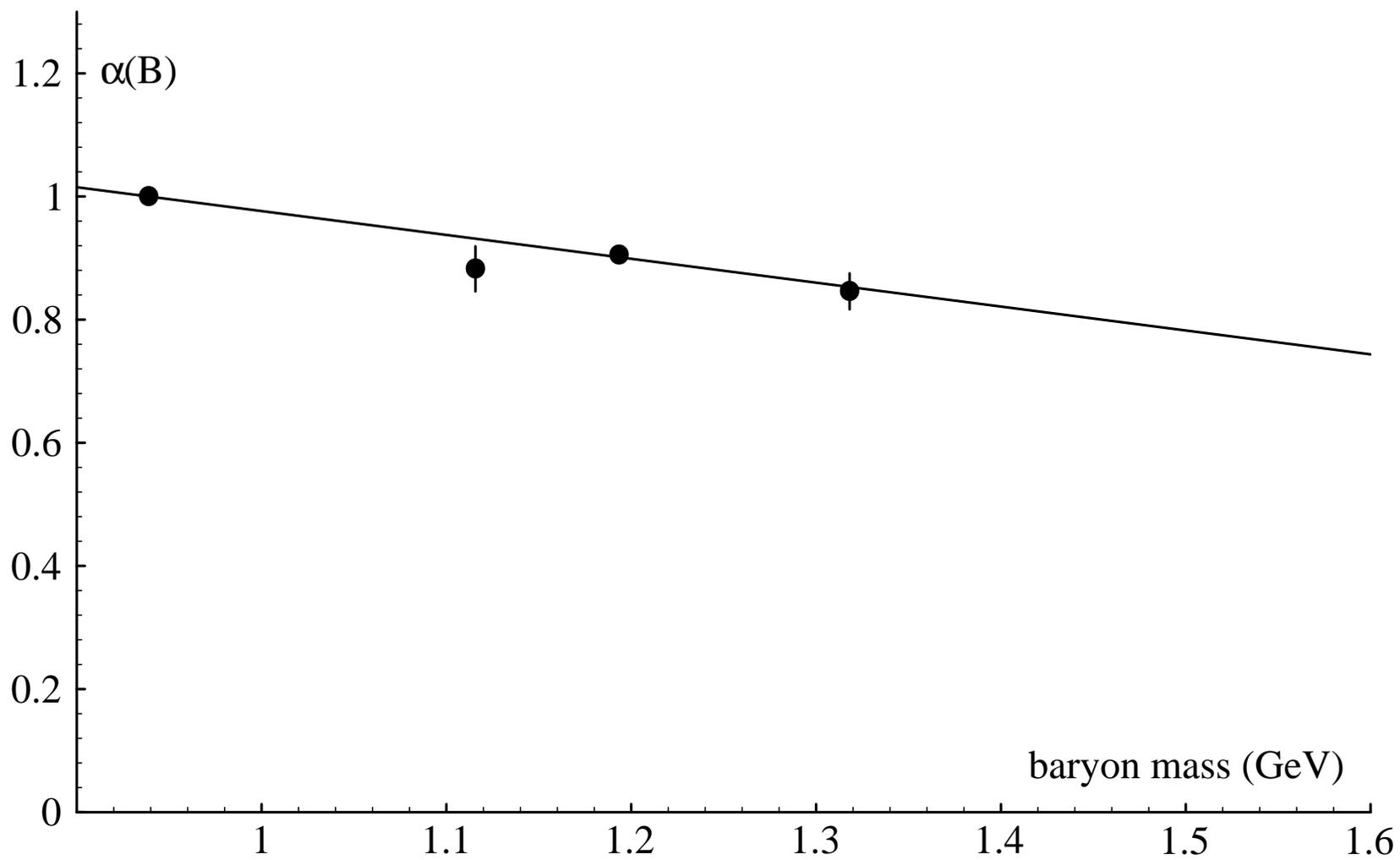
\begin{figure}
\caption{The ratio $\alpha(B)=\mu_d^B/\mu_d^N$ as a function of the 
baryon mass.  The points are the data for the nucleon, $\Lambda$, 
$\Sigma$ and $\Xi$ as given by equations 
(\protect\ref{alphaL})-(\protect\ref{alphaX}).  The straight line 
represents the linear fit according to equation (\protect\ref{lin}).}
\label{cool_plot}
\end{figure}
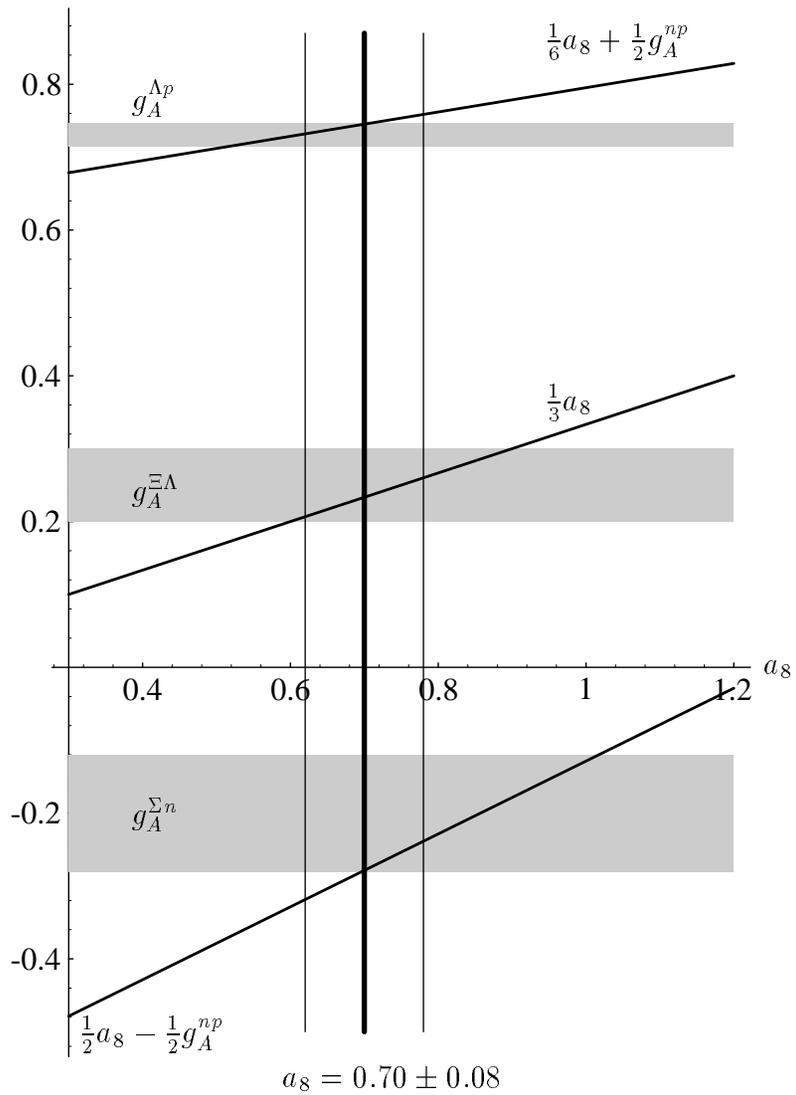
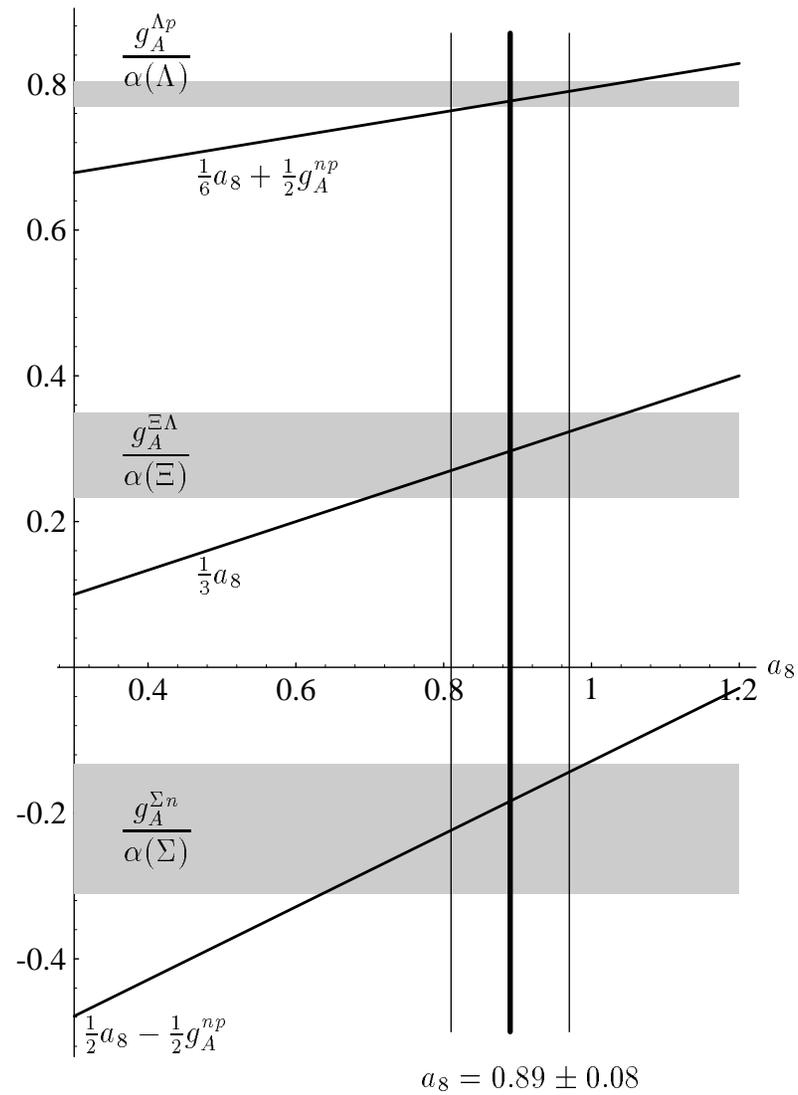
\begin{figure}
	\caption{The left picture illustrates the fit to Eqs.\ 
	(\protect\ref{anpassa1})-(\protect\ref{anpassa3}), without the $\alpha$ 
	factors. The gray bands are the experimental values of the $g_{A}$'s 
	including the errors. The linear functions are the corresponding 
	right hand sides of 
	(\protect\ref{anpassa1})-(\protect\ref{anpassa3}). The thick central 
	vertical line is the value for $a_{8}$ obtained in the fit, and the 
	thinner vertical lines are the error bounds of the 
	fit. The right picture similarly illustrates the fit when the 
	$\alpha$'s are included. We can see that all $g_{A}$'s are better 
	accounted for when using the $\alpha$ factors.}
	\label{cooler_plot}
\end{figure}
\end{document}